\documentclass[twocolumn,english,aps,groupedaddress,showpacs]{revtex4}
\usepackage[T1]{fontenc}
\usepackage[latin9]{inputenc}
\usepackage{amsmath}
\usepackage{graphicx}
\usepackage{amssymb}
\usepackage{esint}

\makeatletter
\@ifundefined{textcolor}{}
{%
 \definecolor{BLACK}{gray}{0}
 \definecolor{WHITE}{gray}{1}
 \definecolor{RED}{rgb}{1,0,0}
 \definecolor{GREEN}{rgb}{0,1,0}
 \definecolor{BLUE}{rgb}{0,0,1}
 \definecolor{CYAN}{cmyk}{1,0,0,0}
 \definecolor{MAGENTA}{cmyk}{0,1,0,0}
 \definecolor{YELLOW}{cmyk}{0,0,1,0}
 }

\usepackage{amsfonts}
\usepackage{longtable}
\usepackage{refcount}

\@ifundefined{lyxdeleted}{}
{}

\makeatother

\usepackage{babel}

\begin{document}

\title{Incoherent topological defect recombination dynamics in TbTe$_{3}$}

\author{T. Mertelj$^{1}$, P. Kusar$^{1}$, V. V. Kabanov$^{1}$, I.Fisher$^{2,3}$
and D. Mihailovic$^{1,4}$}

\affiliation{$^{1}$Complex Matter Department, Jozef Stefan Institute, Jamova
39, 1000 Ljubljana, Slovenia}

\affiliation{$^{2}$Geballe Laboratory for Advanced Materials and Department of
Applied Physics, Stanford University, Stanford, California 94305,
USA}

\affiliation{$^{3}$Stanford Institute for Materials and Energy Sciences, SLAC
National Accelerator Laboratory, 2575 Sand Hill Road, Menlo Park,
California 94025, USA}

\affiliation{$^{4}$CENN Nanocentre, Jamova 39, 1000 Ljubljana, Slovenia}

\date{\today}
\begin{abstract}
We study the incoherent recombination of topological defects created
during a rapid quench of a charge-density-wave system through the
electronic ordering transition. Using a specially devised 3-pulse
femtosecond optical spectroscopy technique we follow the evolution
of the order parameter over a wide range of timescales. By careful
consideration of thermal processes we can clearly identify intrinsic
topological defect annihilation processes on a timescale $\sim30$
ps and find a signature of extrinsic defect-dominated relaxation dynamics
occurring on longer timescales. 
\end{abstract}

\pacs{71.45.Lr, 78.47.jh, 63.20.kp}

\maketitle
Topological defects are non-linear objects which can be created any
time a symmetry-breaking transition occurs.\citep{bunkovGodfrin2000,Kibble1976,Zurek1985}
They can be described theoretically as solutions to systems of nonlinear
differential equations based on Ginzbug-Landau theory. They are of
great fundamental importance in fields such as cosmology where they
appear as strings and condensed matter physics where they appear in
the form of vortices and domain walls. While a good understanding
of static properties of topological defects (TD) has come from systems
such as liquid crystals, the dynamics of TDs are much less understood.
Electronic phase transitions in charge-density wave systems\citep{gruner1994}
are particularly interesting model systems for studying the general
behavior of the dynamics of topological excitations. The collective
excitations are not overdamped which allows the observation of both
collective and quasiparticle (QP) excitations as they evolve through
the transition. In particular, they can be used to investigate the
dynamic behavior of topological excitations such as domain walls in
real time using ultrafast laser techniques. 

Recently, time-resolved experiments have shown that following a quench
created by a strong laser pulse the order parameter (OP) oscillates
coherently, revealing coherent TD dynamics.\citep{YusupovMertelj2010}
Domain walls are created parallel to the crystal surface which can
coherently annihilate on the timescale of a few picoseconds with the
accompanying emission of collective modes which have been detected
as modulations of reflectivity upon reaching the surface. In addition
to coherent defect dynamics, incoherent topological defects created
by the Kibble-Zurek mechanism\citep{Kibble1976} are also expected,
but very little is known about the dynamics of incoherent TD dynamics
in CDW systems, and in condensed matter systems in general. 

In this paper we investigate the incoherent evolution of TDs using
a specially devised 3-pulse femtosecond spectroscopy\citep{YusupovMertelj2010}
technique which allows the direct background free observation of the
evolution of the order parameter (OP) as a function of time through
the electronic ordering transition. In a rapid quench experiment order
emerges in different regions of the sample independently so multiple
topological defects can be created. Their presence can be detected
in the optical response as a spatial inhomogeneity of the order parameter.
The determination of incoherent TD dynamics is a challenging task,
however. Because of thermal diffusion processes, which evolve on similar
timescales as topological annihilation and also introduce temperature
inhomogeneity, careful temperature calibration from independently
measured frequencies is needed to accurately account for thermal effects.
We deal with the problem by careful calibration of the transient effective
temperatures, which enables us to unambiguously distinguish the incoherent
dynamics from thermal diffusion effects. 

\begin{figure}[h]
\begin{centering}
\includegraphics[width=0.9\columnwidth]{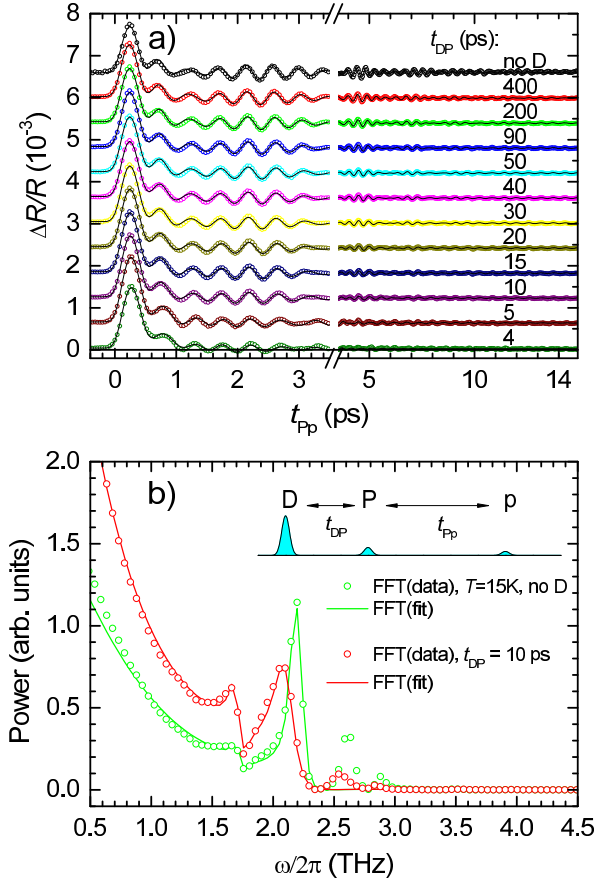} 
\par\end{centering}

\caption{\label{fig:transients}a) The transient reflectivity of TbTe$_{3}$
at different times $t_{\mathrm{DP}}$ after the D pulse. The thin
lines are the fits discussed in text. b) An example of the FFT power
spectra for the raw data and the fit at $t_{\mathrm{DP}}=10$ ps with
and without a D pulse. The inset shows the laser pulse sequence.}

\end{figure}

In our experiments, we use a three pulse technique described in refs.
{[}\onlinecite{YusupovMertelj2010,KusarMertelj2011}{]}: A \textquotedbl{}destruction\textquotedbl{}
(D) laser pulse at 800 nm excites a cold sample%
\footnote{The growth of the samples by a self-flux technique and subsequent
characterization was described elsewhere.\citep{RuCondron2008}%
} into the disordered state, breaking up the CDW order. We then monitor
the evolution of the transient reflectivity $\Delta R(t_{\mathrm{Pp}})/R$
excited with a weaker pump (P) pulse as a function of time delay $t_{\mathrm{DP}}$
between the D and P pulse (the pulse sequence nomenclature is illustrated
in the insert to Fig. 1b)). The D pulse fluence is adjusted to twice
the threshold for causing the destruction of the ordered state \citep{YusupovMertelj2010}.
After a quench by the laser pulse, order recovers first through the
sub-picosecond recovery of the quasiparticle gap leading to coherent
oscillations of the OP and the coherent creation of TDs which decay
within 5-8 ps in TbTe$_{3}$ \citep{YusupovMertelj2010}. Since the
CDW coherence length ($\sim$2 nm) is much shorter than the size of
the laser excited volume ($\sim50$$\mu$m dia), order emerges with
different phase in different regions, resulting in the formation of
topological defects whose spatial distribution is determined partly
by the inhomogeneous excitation and partly by the underlying fluctuations
which nucleate the emergence of order by the so called Kibble-Zurek\citep{Kibble1976,Zurek1985}
mechanism. The resulting inhomogeneity of the OP leads to observable
temporally resolvable effects in the frequency, linewidth and amplitude
of the collective amplitude mode, all of which are related to the
OP, as shown in previous studies\citep{YusupovMertelj2008}.

\begin{figure}[h]
\begin{centering}
\includegraphics[clip,width=1\columnwidth]{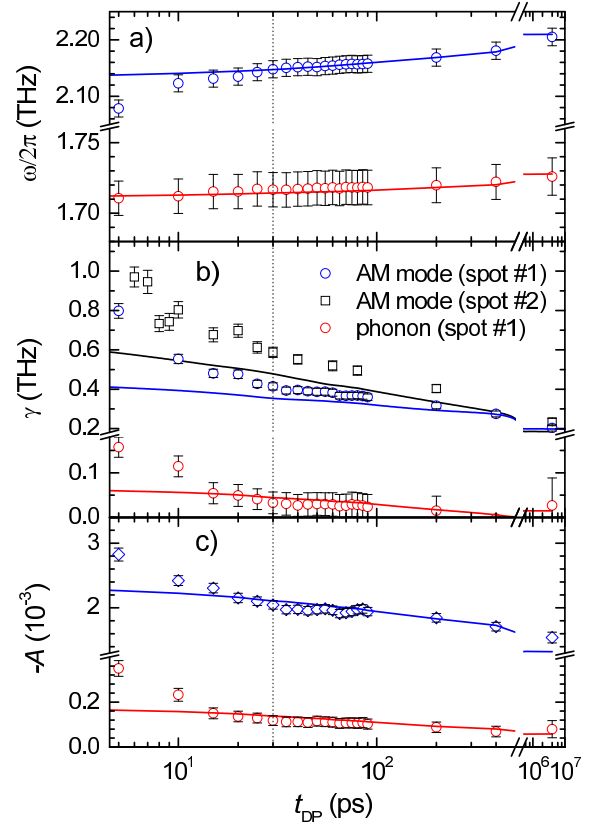} 
\par\end{centering}

\caption{\label{fig:AvstDP}a) Frequencies of the AM and the 1.7-THz phonon
as a function of $t_{\mathrm{DP}}$. b) Effective dampings of the
AM and the phonon as a function of $t_{\mathrm{DP}}$. Open squares
represent a measurement from another spot on the sample. c) Amplitudes
of the AM and the phonon as a function of $t_{\mathrm{DP}}$. The
solid lines are the frequencies, linewidths and amplitudes calculated
using an inhomogeneous temperature distribution model.\citep{suppl}
Since no special care was taken to calibrate the D-pulse beam diameter
at the sample for the measurement on spot \#2 the actual D-pulse fluence
on spot \#2 was slightly higher leading to different TDM parameters.}

\end{figure}

In Fig. \ref{fig:transients}(a) we show the raw data on the transient
reflectivity $\Delta R(t_{\mathrm{Pp}})/R$ of TbTe$_{3}$ at different
time delays $t_{\mathrm{DP}}$ after the $D$ pulse. After the initial
QP relaxation we observe oscillations due to the coherently excited
order parameter amplitude mode (AM) and other phonons.\citep{SchmittKirchmann2008,YusupovMertelj2008,LavagniniEiter2010}
The level of noise is very small, due to the the excellent intrinsic
properties of the material, which helps us make a detailed quantitative
analysis. We analyze the transient reflectivity oscillations using
the theory for displacive excitation of coherent phonons:\citep{ZeigerVidal1992}\begin{multline}
\frac{\Delta R(t_{\mathrm{Pp}})}{R}=\int_{0}^{\infty}G(t-u)\times\\
\qquad\times[A_{\mathrm{e}}\exp(-u/\tau)+A_{\mathrm{B}}]du+\\
+\sum A_{i}\int_{0}^{\infty}G(t-u)\exp(-\gamma_{i}u)\times\\
\times[\cos(\Omega_{i}u)-\beta_{i}\sin(\Omega_{i}u)]du\label{eq:displ}\end{multline}
where $\beta_{i}=(1/\tau-\gamma_{i})/\Omega_{i}$, $G(t)=\exp(-2t^{2}/\tau_{\mathrm{P}}^{2})$
and $\tau_{\mathrm{P}}$ is the laser pulse length. The first integral
represents the QP relaxation with the relaxation time $\tau$, while
the sum depicts the response of coherent phonons with frequencies
$\Omega_{i}$, and effective dampings $\gamma_{i}$. $A_{\mathrm{e}}$
corresponds to the QP relaxation amplitude while the residual value
at long delays is $A_{\mathrm{B}}$. To limit the number of fitting
parameters we keep only two phonon terms corresponding to the AM at
2.2 THz and the 1.7-THz phonon which strongly interacts with the AM
at higher temperatures.\citep{YusupovMertelj2008} Fig \ref{fig:transients}(b)
shows the fast Fourier transform (FFT) of the raw data and of the
fit to the data i) without the D pulse and ii) for a D-P delay of
$t_{\mathrm{DP}}=10$ ps, clearly showing that Eq. (\ref{eq:displ})
fits the response very well below $\sim2.4$ THz irrespective of $t_{\mathrm{DP}}$.
The $t_{\mathrm{DP}}$-dependence of the frequency, linewidth and
amplitude of the AM and 1.7 THz phonon modes are shown in Fig. \ref{fig:AvstDP}.
The linewidth is shown for two sets of data obtained from different
spots on the sample in two separate measurements.

\begin{figure}[h]
\begin{centering}
\includegraphics[clip,width=1\columnwidth]{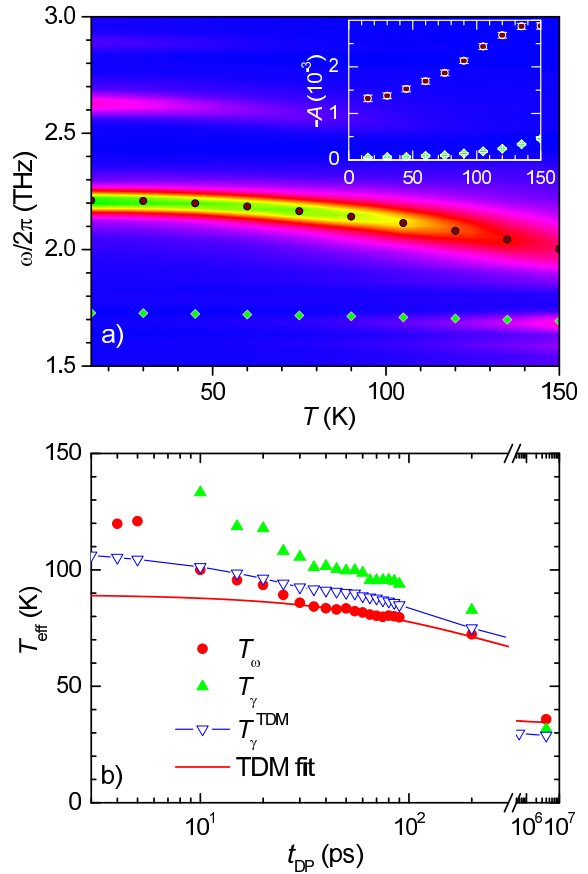} 
\par\end{centering}

\caption{\label{fig:Teff}a) The temperature dependence of the coherent oscillations
spectrum measured in a standard Pp experiment. Circles and diamonds
represent the frequencies obtained from the time domain fit for the
AM and the 1.7-THz phonon, respectively. The inset shows the $T-$dependence
of the amplitudes for both modes. b) The time-dependence of the effective
temperature $T_{\mathrm{eff}}(t_{\mathrm{DP}})$ from $\omega_{\mathrm{AM}}$($t_{\mathrm{DP}})$
and $\gamma_{\mathrm{AM}}(t_{\mathrm{DP}})$. The solid line is a
fit to $\omega_{\mathrm{AM}}$ effective temperature using the thermal
diffusion model. The inverted triangles correspond to the calculated
$T_{\gamma}^{\mathrm{TDM}}(t_{\mathrm{DP}})$ taking into account
the inhomogeneous temperature distribution.\citep{suppl}}

\end{figure}

In order to obtain a calibration of the effective temperature $T_{\mathrm{eff}}$
of the photoexcited sample volume we measured \emph{independently},
by means of a standard pump-probe experiment, the $T$-dependence
of the reflectivity transients in the thermal equilibrium and determine
the $T$-dependent amplitude, frequency {[}$\omega_{\mathrm{AM}}(T)${]}
and damping {[}$\mathrm{\gamma}_{\mathrm{AM}}(T)${]} for the AM.
Using these calibrations we are in a position to determine $T_{\mathrm{eff}}$
as a function of time from $\omega_{\mathrm{AM}}(t_{\mathrm{DP}})$
and $\gamma_{\mathrm{AM}}(t_{\mathrm{DP}})$ and take it into account
to obtain the thermal inhomogeneity dynamics. The time-dependence
of the $T_{\mathrm{eff}}$ is shown in Fig. \ref{fig:Teff}(b). We
observe that the two effective temperatures $T_{\omega}(t_{\mathrm{DP}})$
and $T_{\gamma}(t_{\mathrm{DP}})$ obtained from the $\omega_{\mathrm{AM}}(t_{\mathrm{DP}})$
and $\gamma_{\mathrm{AM}}(t_{\mathrm{DP}})$ systematically differ
by approximately 20 K indicating an excess AM linewidth with respect
to the thermal equilibrium state.%

One of the most obvious contributions to the excess AM linewidth is
the inhomogeneous broadening caused by the thermal inhomogeneity.
In order to be able to determine and analyze any other contributions
to the linewidth we therefore determine the contribution of the thermal
inhomogeneity to the excess AM linewidth. To do this we first fit
a thermal diffusion model\citep{MerteljOslak2009,suppl} (TDM) to
the effective temperature obtained from the AM frequency $T_{\omega}(t_{\mathrm{DP}})$,
and then use the TDM parameters to calculate the transient optical
 reflectivity which fully takes into account inhomogeneity of the
temperature in the excited volume.\citep{suppl} As seen in Fig. \ref{fig:Teff}
(b), $T_{\omega}(t_{\mathrm{DP}})$ can be fit very well over 5 decades
of time from 30 ps to 4 $\mu$s using a one-dimensional%
\footnote{The diameters of the beams are much larger that the optical penetration
depth so on the relevant timescale the heat diffusion is 1D.%
} TDM, where $T(t_{\mathrm{DP}})=\Delta T/\sqrt{1+t_{\mathrm{DP}}/\tau_{\mathrm{D}}}+T_{0}$,
and the fit parameter $\tau_{\mathrm{D}}\sim120$ ps represents the
characteristic heat diffusion time\citep{suppl}.

Comparing now the simulation\citep{suppl} with the experiment in
Fig. \ref{fig:AvstDP}, we see that the validity of the TDM beyond
$\sim30$ ps is well supported by the good agreement of the simulation
for both the AM and the 1.7-THz phonon parameters. We can thus safely
conclude, that the recovery of the order parameter on timescales longer
than $\sim30$ ps is primarily governed by the 1D heat diffusion process. 

\begin{figure}
\includegraphics[angle=-90,width=0.9\columnwidth]{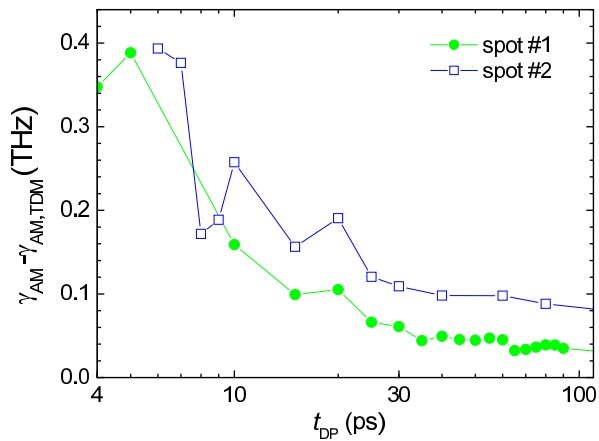}

\caption{The time-dependence of the excess $\gamma_{\mathrm{AM}}$ due to topological
defect annihilation. Data for two different spots on the sample show
a difference primarily in the long time behavior\label{fig:diff-vs-tDP}
beyond 30 ps, which appears as an offset.\citep{mainfoot}}

\end{figure}

Below $\sim30$ ps however, there is a large discrepancy between the
calculated $T_{\mathrm{eff}}$, $\gamma_{\mathrm{AM}}$ and other
phonon parameters in comparison to the data, even after carefully
taking into account the thermal inhomogeneity. The observed magnitude
and the evolution of $\gamma_{\mathrm{AM}}(t_{\mathrm{DP}})$ for
$t_{\mathrm{DP}}$$<30$ ps clearly \textit{\emph{cannot be assigned
}}solely to the temperature inhomogeneity\emph{.} Subtracting the
thermal inhomogeneity contribution from the AM linewidth in Fig. 2
b), we can now isolate the topological-defects inhomogeneity contribution
 as shown in Fig. \ref{fig:diff-vs-tDP}.

As discussed in the introduction some of the defects created in the
quench process annihilate \emph{coherently} resulting in an aperiodic
modulation of the AM intensity and frequency in the first $\sim$8
picoseconds\citep{YusupovMertelj2010}. The very low level of experimental
noise in the raw data allows us to attribute the observed data scatter
in Fig. \ref{fig:diff-vs-tDP} to the coherent defect dynamics in
the material, rather than experimental noise. 

The modulation of the AM due to the defects that annihilate \emph{incoherently}
is unfortunately not detected \emph{directly} by our stroboscopic
technique. However, the incoherent topological defects give rise to
a spatial inhomogeneity of the order parameter and a\emph{ decoherence}
of the AM oscillations leading to an increased linewidth $\gamma_{\mathrm{AM}}$
for $t_{\mathrm{DP}}<30$ ps which we have detected in our experiments.
Concurrently the defects give rise to a softening of the collective
mode $\omega_{\mathrm{AM}}$, because of the OP suppression which
they cause. The increase of the coupled 1.7-THz phonon effective damping
at shorter $t_{\mathrm{DP}}$ as shown in Fig. \ref{fig:AvstDP} is
presumably also caused by the inhomogeneity of the OP. A further manifestation
of the incoherent annihilation is the increase of the amplitudes of
the AM and the phonon with respect to the TDM for $t_{\mathrm{DP}}<\sim30$
ps, which is also consistent with the suppression of the OP when one
takes into account that the amplitudes of the modes increase with
the decreasing OP amplitude deduced from their $T$-dependence shown
in the inset of Fig. \ref{fig:Teff}(a).

Apart from intrinsic topological defect annihilation processes, which
we have identified on a timescale of $\sim$30 ps, we expect to observe
annihilation of domain walls pinned to defects and imperfections at
longer times. The timescale of their annihilation may extend well
beyond 30 ps. Evidence for such slower extrinsic recombination processes
comes from the  long time behavior shown in Fig. 4. $\gamma_{\mathrm{AM}}$
remains systematically larger than the predicted thermally inhomogeneous
linewidth $\gamma_{\mathrm{AM}}^{\mathrm{TDM}}$ suggesting a slower
recombination of the pinned domain walls.\citep{mainfoot} 

Recently the absence of topological defects on ultrafast timescales
in highly excited charge ordered (CO) nickelate was suggested.\citep{LeeChuang2012}
A slow relaxation of the CO X-ray diffraction peak intensity, on the
timescale of $\sim60$ ps, was, due to the absence of any increase
of the diffraction peak linewidth, attributed to a depopulation of
the phason mode. While, contrary to \citet{LeeChuang2012}, our excitation
density is clearly high enough to excite topological defects,\citep{YusupovMertelj2010}
there exists a possible anharmonic contribution of the highly excited
phason mode to the AM linewidth. The anharmonic processes, however,
contribute to both, the linewidth and frequency renormalization of
the AM,\citep{Balkanski1983} and \emph{can not} lead to the difference
between $T_{\omega}$ and $T_{\gamma}$ as observed in the experiment.

In conclusion, these experiments demonstrate the possibility of studying
both coherent and incoherent topological defects dynamics in complex
materials in which the order parameter can be monitored in real time
through the dynamics of the collective mode. The dynamics on a timescale
of $\sim30$ ps can unambiguously be associated with intrinsic topological
defects annihilation in TbTe$_{3}$ following a laser quench arising
from the time-dependent inhomogeneity and suppression of the order
parameter. The inhomogeneity causes an increased effective damping
of the amplitude mode while the suppression of the order-parameter
is indicated by an additional softening of the AM-mode frequency.
Beyond $\sim30$ ps we find a predominantly thermal-diffusion governed
order-parameter dynamics with a signature of extrinsic defect annihilation
dynamics. 

\bibliographystyle{apsrev}
\bibliography{biblio}

\begin{thebibliography}{16}
\bibitem[{mai()}]{mainfoot}
\bibinfo{note}{In the standard 2-pulse pump-probe experiments we observe some
  spatial variation of the AM linewidth which we assign to the intrinsic
  structural-defect/impurity density spatial variation in the sample. Since, in
  the absence of 2-pulse $T$-dependent data for spot \#2, the TDM simulation
  for spot \#2 was based on $T$-dependent scans from spot \#1 the offset
  between the spots can be attributed to the systematic error due to the
  spatial variation of the AM linewidth.}
\end{thebibliography}
\begin{thebibliography}{2}
\bibitem[{sup()}]{supplfoot}
\bibinfo{note}{In the standard 2-pulse pump-probe experiments we observe some
  spatial variation of the AM linewidth which we assign to the intrinsic
  structural-defect/impurity density spatial variation in the sample. In the
  absence of 2-pulse $T$-dependent data for spot \#2 the TDM simulation for
  spot \#2 was based on $T$-dependent scans from spot \#1. This lead to the
  systematic error due to the spatial variation of the AM linewidth.}
\end{thebibliography}

\end{document}

\title{Incoherent topological defect recombination dynamics in TbTe$_{3}$:
supplemental information}

\author{T. Mertelj$^{1}$, P. Kusar$^{1}$, V. V. Kabanov$^{1}$, I.Fisher$^{2,3}$
and D. Mihailovic$^{1,4}$}

\affiliation{$^{1}$Complex Matter Department, Jozef Stefan Institute, Jamova
39, 1000 Ljubljana, Slovenia}

\affiliation{$^{2}$Geballe Laboratory for Advanced Materials and Department of
Applied Physics, Stanford University, Stanford, California 94305,
USA}

\affiliation{$^{3}$Stanford Institute for Materials and Energy Sciences, SLAC
National Accelerator Laboratory, 2575 Sand Hill Road, Menlo Park,
California 94025, USA}

\affiliation{$^{4}$CENN Nanocentre, Jamova 39, 1000 Ljubljana, Slovenia}

\date{\today}

\maketitle

\section{Thermal diffusion model simulation}

\subsection{Temperature inhomogeneity}

On longer time scales, when the system is locally thermalized, it
can be described by the microscopical thermodynamic temperature, which
is governed by the heat-diffusion equation, $\frac{\partial T}{\partial t}=D(T)\nabla^{2}T$.
Due to the experimental geometry, where the experimental volume has
a shape of a high aspect ratio pancake, with the diameter much larger
than the thickness, the diffusion is one dimensional (1D) on all relevant
timescales.

The 1D solution of the heat-diffusion equation, using the von Neumann
boundary condition, $\frac{\partial T(t,z)}{\partial z}=0$, at $z=0$
and assuming a $T$ independent diffusion constant $D$, is: \begin{equation}
\Delta T=\frac{1}{\sqrt{4\pi Dt}}\int_{0}^{\infty}\left(e^{\frac{-(z-u)^{2}}{4Dt}}+e^{\frac{-(z+u)^{2}}{4Dt}}\right)\Delta T(0,u)du.\end{equation}
When the initial temperature $\Delta T(0,z)$ is given by the exponentially
decaying laser pulse depth profile, $\Delta T(0,z)=\Delta T\exp\left(-z/\lambda_{\mathrm{p}}\right)$,
with $\lambda_{p}$ being the optical penetration depth in units of
the correlation length and $\Delta T$ a positive coefficient depending
on the intensity of the pulse, we obtain the solution: \begin{align}
\Delta T(t,z) & =\frac{\Delta T}{2}e{}^{t/4\tau_{\mathrm{D}}}\cdot\nonumber \\
 & \cdot\left[e^{-z/\lambda_{\mathrm{p}}}\text{erfc}\left(\sqrt{\frac{t}{4\tau_{\mathrm{D}}}}-\frac{z}{\lambda_{\mathrm{p}}}\sqrt{\frac{\tau_{\mathrm{D}}}{t}}\right)\right.+\nonumber \\
 & +\left.e^{z/\lambda_{\mathrm{p}}}\text{erfc}\left(\sqrt{\frac{t}{4\tau_{\mathrm{D}}}}+\frac{z}{\lambda_{\mathrm{p}}}\sqrt{\frac{\tau_{\mathrm{D}}}{t}}\right)\right].\label{eq:sol-exact}\end{align}
Here $\tau_{\mathrm{D}}=\lambda_{\mathrm{p}}^{2}/4D$ is the characteristic
heat diffusion time. A simpler solution is found using the initial
temperature $\Delta T(0,z)$ with a Gaussian profile\citep{MerteljOslak2009}
$\Delta T_{\mathrm{G}}(0,z)=\Delta T\exp\left(-z^{2}/\lambda_{p}^{2}\right),$
\begin{equation}
\Delta T_{\mathrm{G}}(t,z)=\frac{\Delta T}{\sqrt{\left(1+t/\tau_{\mathrm{D}}\right)}}\exp\left(-\frac{z^{2}}{\lambda_{p}^{2}\left(1+t/\tau_{\mathrm{D}}\right)}\right).\label{eq:solheat1}\end{equation}

\subsection{Optical reflectivity transients in the presence of the temperature
inhomogeneity}

\begin{figure}
\includegraphics[clip,angle=-90,width=1\columnwidth]{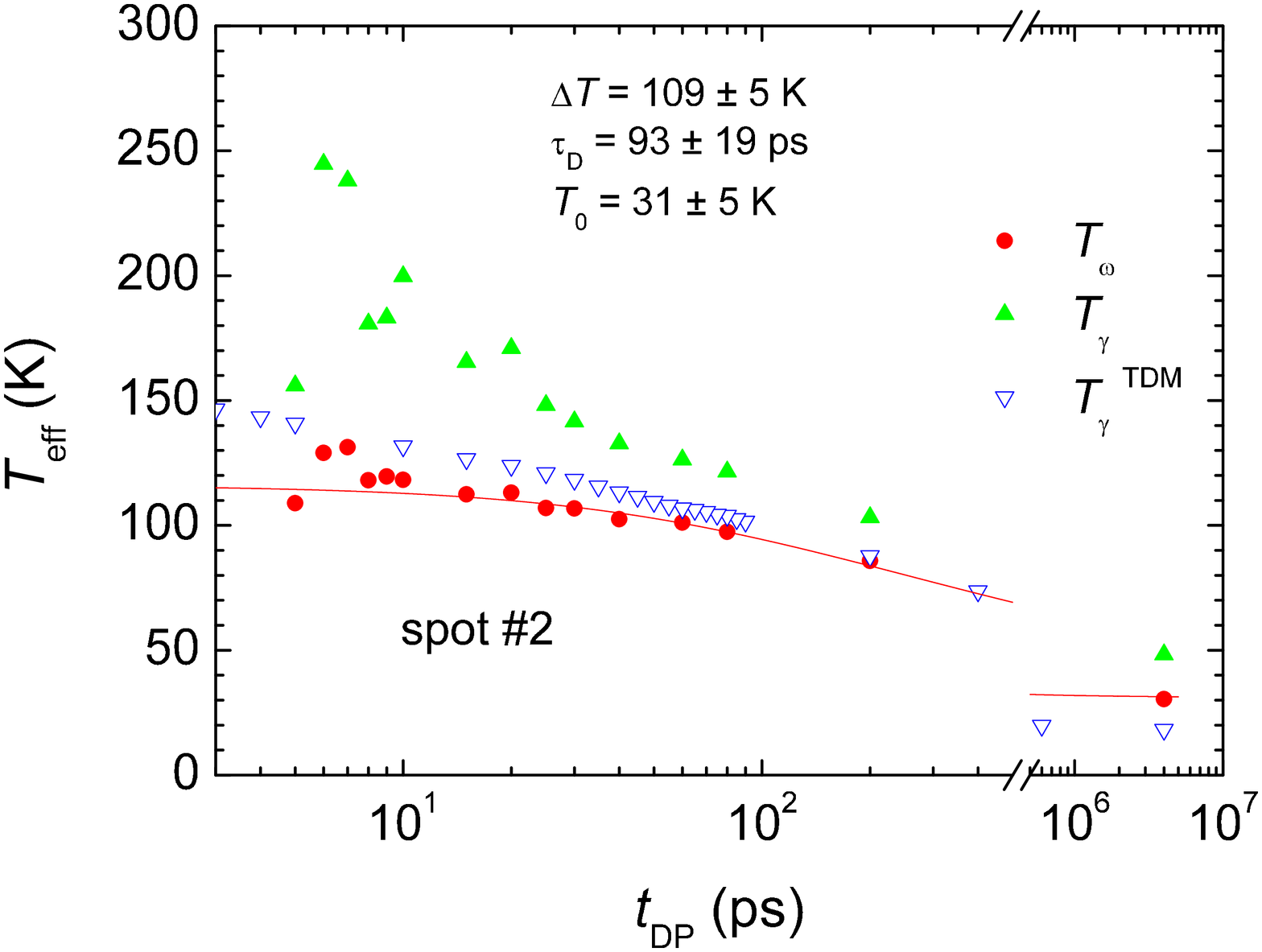}

\caption{The time-dependence of the effective temperature $T_{\mathrm{eff}}(t_{\mathrm{DP}})$
obtained from $\omega_{\mathrm{AM}}$($t_{\mathrm{DP}})$ and $\gamma_{\mathrm{AM}}(t_{\mathrm{DP}})$
at spot \#2. The solid line is a fit to $\omega_{\mathrm{AM}}$ effective
temperature using the thermal diffusion model (\ref{eq:solheat1}).
The inverted triangles correspond to the calculated $T_{\gamma}^{\mathrm{TDM}}(t_{\mathrm{DP}})$
taking into account the inhomogeneous temperature distribution. The
underestimation of $T_{\gamma}^{\mathrm{TDM}}$ at long $t_{\mathrm{DP}}$
is due to the intrinsic variation of the thermal equilibrium $\gamma_{\mathrm{AM}}$.\citep{supplfoot}}

\label{fig:Teff}
\end{figure}

To calculate the excess AM linewidth due to the inhomogeneous broadening
caused by the thermal inhomogeneity we first fit a thermal diffusion
model (\ref{eq:solheat1}) (TDM) to the effective temperature obtained
from the AM frequency, $T_{\omega}(t_{\mathrm{DP}})$. As seen in
Fig. 3(b) and \ref{fig:Teff}, $T_{\omega}(t_{\mathrm{DP}})$ can
be fit very well over 5 decades of time from 30 ps to 4 $\mu$s.

 Next we simulate the optical reflectivity response in the presence
of the inhomogeneous temperature distribution using the thermal diffusion
parameters obtained from the fit above. The response is given by:\begin{eqnarray}
\left.\frac{\Delta R\mathrm{_{TDM}}(t_{\mathrm{Pp}})}{R}\right|_{t\mathrm{_{DP}}} & = & \int_{0}^{\infty}e^{-3z/\lambda_{\mathrm{op}}}\times\nonumber \\
 &  & \qquad\times\left.\frac{\Delta R(t_{\mathrm{Pp}})}{R}\right|_{T(z,t_{\mathrm{DP}})}dz,\label{eq:integral}\\
T(z,t_{\mathrm{DP}}) & = & \Delta T(t_{\mathrm{DP}},z)+T_{0}.\qquad\end{eqnarray}
Here $\lambda_{\mathrm{op}}$ is the optical penetration depth and
$\Delta R(t_{\mathrm{Pp}})/R|_{T(t_{\mathrm{DP}},z)}$ in (\ref{eq:integral})
is calculated by interpolation from the $T$-dependent transients
which were independently measured in the thermal equilibrium. From
the calculated response in Eq. (\ref{eq:integral}) we then determine
the predicted thermally inhomogeneous linewidth $\gamma_{\mathrm{AM}}^{\mathrm{TDM}}(t_{DP})$,
and other phonon parameters, as well as the effective temperature
$T_{\gamma}^{\mathrm{TDM}}(t_{\mathrm{DP}})$ in the same way as previously,
by fitting Eq. (1).

\bibliographystyle{apsrev}
\bibliography{biblio}